\documentclass[sigconf]{acmart}
\usepackage{booktabs}

\AtBeginDocument{%
  }

\setcopyright{acmlicensed}
\copyrightyear{2018}
\acmYear{2018}
\acmDOI{XXXXXXX.XXXXXXX}

\copyrightyear{2025}
\acmYear{2025}
\setcopyright{cc}
\setcctype{by-nc-sa}
\acmConference[FAccT '25]{The 2025 ACM Conference on Fairness,
Accountability, and Transparency}{June 23--26, 2025}{Athens, Greece}
\acmBooktitle{The 2025 ACM Conference on Fairness, Accountability, and
Transparency (FAccT '25), June 23--26, 2025, Athens,
Greece}\acmDOI{10.1145/3715275.3732007}
\acmISBN{979-8-4007-1482-5/2025/06}

\begin{document}

\title[From Efficiency Gains to Rebound Effects]{From Efficiency Gains to Rebound Effects: The Problem of Jevons' Paradox in AI's Polarized Environmental Debate}

\author{Alexandra Sasha Luccioni}
\email{sasha.luccioni@hf.co}

\affiliation{%
  \institution{Hugging Face}
  \city{Montreal}
  \country{Canada}
}

\author{Emma Strubell}
\affiliation{%
  \institution{Carnegie Mellon University}
  \city{Pittsburgh}
  \country{USA}}

\author{Kate Crawford}
\affiliation{%
  \institution{Microsoft Research; University of Southern California}
  \city{New York}
  \country{USA}
}

\renewcommand{\shortauthors}{Luccioni et al.}






\begin{abstract}
    
As the climate crisis deepens, artificial intelligence (AI) has emerged as a contested force: some champion its potential to advance renewable energy, materials discovery, and large-scale emissions monitoring, while others underscore its growing carbon footprint, water consumption, and material resource demands. Much of this debate has concentrated on direct impacts—energy and water usage in data centers, e-waste from frequent hardware upgrades—without addressing the significant indirect effects. This paper examines how the problem of Jevons’ Paradox applies to AI, whereby efficiency gains may paradoxically spur increased consumption. We argue that understanding these second-order impacts requires an interdisciplinary approach, combining lifecycle assessments with socio-economic analyses. Rebound effects undermine the assumption that improved technical efficiency alone will ensure net reductions in environmental harm. Instead, the trajectory of AI’s impact also hinges on business incentives and market logics, governance and policymaking, and broader social and cultural norms. We contend that a narrow focus on direct emissions misrepresents AI’s true climate footprint, limiting the scope for meaningful interventions. We conclude with recommendations that address rebound effects and challenge the market-driven imperatives fueling uncontrolled AI growth. By broadening the analysis to include both direct and indirect consequences, we aim to inform a more comprehensive, evidence-based dialogue on AI’s role in the climate crisis.

\end{abstract}

\maketitle

\section{Introduction} \label{sec:intro}

As the climate crisis intensifies, the environmental impacts of artificial intelligence (AI) have become the subject of many a polarized debate. The question of whether the potential positive impacts of AI outweigh the negative ones, and how to foster technological advances in AI while minimizing its environmental harms, has divided AI researchers and practitioners alike. Some maintain that its potential to accelerate sustainable breakthroughs will exceed its environmental costs  by increasing renewable energy production and transmission or aiding in the design of more sustainable materials \cite{rolnick2022tackling,merchant2023scaling}.  Others point to the soaring resource demands of large-scale AI models and their negative environmental impacts from non-renewable energy use, water consumption, and extraction of minerals~\cite{IEA2024,guidi2024environmental,crawford2021atlas,dauvergne2022artificial}. These opposing positions tend to center on the technology’s direct impacts, measured in energy consumption and greenhouse gas (GHG) emissions from data centers, or in the e-waste that accumulates as hardware becomes obsolete.

Yet a critical dimension of AI’s climate footprint lies outside these direct resource and emissions calculations. Recent work on indirect impacts  \cite{kaack2022aligning} warns of potential “rebound effects”, whereby gains in efficiency spur higher overall consumption. These second-order effects challenge the presumption that purely technical optimizations alone will deliver sufficient climate benefits. Cost savings achieved by more efficient AI hardware, for example, can spur increased demand for new AI functionalities, which in turn drive further hardware upgrades and increase costs. Economists refer to such transformations as \textit{Jevons’ Paradox}, which was proposed in the 19th century by economist William Stanley Jevons, who observed that as coal use became more efficient, it was also paradoxically leading to an increase, and not a decrease, in the consumption of coal across different industries~\cite{jevons1866coal}. 
The addition of increasingly efficient AI to systems from commerce to transportation can have far-reaching effects on our societies, our behaviors, and the future paths available to us in the race against climate change. 

This system-level complexity underscores the inadequacy of the question, ``Is AI net positive or net negative for the climate?'' Instead, we adopt an analytic approach that includes the intersecting social, political, and economic contexts in which AI systems are developed and deployed. Although efficiency has been a defining ethos in recent AI research, reflected in ``scaling laws'' that promise ever-more-powerful models~\cite{kaplan2020scaling}, effective climate action requires grappling with how these systems reshape markets, cultural norms, and policy priorities. Thus, understanding rebound effects requires drawing on both qualitative and quantitative methods, drawn from computer science, economics and the social sciences, as they hinge not on algorithmic design but human adaptation and use patterns. Adopting an interdisciplinary approach allows for both a rigorous lifecycle accounting of direct effects as well as understanding the social behaviors that AI can induce or displace.

This paper aims to bridge the gap in the existing literature by, first, providing a brief overview of the debate about AI's direct positive and negative impacts on the environment (\S\ref{sec:direct}). We follow this with an in-depth exploration of AI’s indirect impacts, including systemic rebound effects, and propose actionable strategies to mitigate them (\S\ref{sec:indirect}). We conclude by proposing promising directions for future exploration and research toward improving our understanding of the full spectrum of AI's impacts on the environment. In short, we argue that meaningfully contending with AI's climate impacts requires grappling with both direct and indirect effects. Otherwise, the industry risks pinning its hopes on technical efficiency gains alone without recognizing the social, cultural, and economic contexts that materially shape technology uses. Our aim is to support the field in moving away from polarized extremes and toward a nuanced position that acknowledges the state of climate change and the pressing need to manage all of AI's climate impacts. By expanding this dialogue and grounding it in evidence across multiple disciplines, we can develop strategies that genuinely address AI's role in environmental sustainability.

\section{AI and the environment} \label{sec:direct}

The environmental impacts of AI were largely overlooked until recent years. Now these issues figure prominently in both scientific debates and public media. Central to this discourse is the question of whether AI's capacity to help mitigate climate change, e.g. by optimizing energy use or discovering sustainable materials, truly exceeds its environmental costs in terms of energy consumption, water usage, and mineral extraction. Some contend that AI's potential benefits justify widespread deployment across various climate-focused applications, whereas others caution that unrestrained expansion may ultimately be more harmful than beneficial. In this section, we review these competing perspectives on AI’s positive and negative environmental impacts as presented in academic and industry discussions.

\subsection{Arguments that AI is a net climate negative}

Research into AI’s total usage of natural resources is still nascent, but initial studies have highlighted significant concerns across multiple domains of direct impact. Papers have addressed carbon emissions from training large models~\cite{strubell2019energy,lacoste2019quantifying,luccioni2022estimating,luccioni2023power}, water consumption for cooling servers~\cite{mytton2021data,hogan2015}, and the mining of minerals for technical infrastructure~\cite{crawford2021atlas}. In the present section, we present the different directions of study pursued in terms of AI's negative environmental impacts and discuss the observations that have been put forward. 

\paragraph{AI is using increasing amounts of energy, most of it generated from nonrenewable sources.} Data from the International Energy Agency (IEA) has provided the strongest international benchmark. They note that the electricity demand from data centers, driven heavily by AI training and inference, is currently at 2\% of total global electricity and will more than double by 2026, surpassing Canada’s national power use~\cite{IEA2023,guidi2024environmental}. This growth is putting strain on energy grids around the world and resulting in many countries, such as Ireland and the Netherlands, placing a moratorium on the construction of new data centers in certain regions. In the United States, the demand for electricity has surged in the past twelve months – according to a recent report into the national grid, electric utilities have nearly doubled their forecasts of how much additional power they will need in the next five years~\cite{wilson2023era}. 

\paragraph{AI data centers are putting stress on already strained aquifers globally.} The construction and operation of data centers requires vast quantities of water, which is used for cooling servers to prevent them from overheating by circulating cool water through radiators. This cooling process requires a constant supply of cool, fresh water, which is heated up during the process, causing a significant portion of it to evaporate, whereas the rest has to be cooled and filtered before being reused or discharged back into local aquifers~\cite{ristic2015water}. Corporate reports have revealed the scale of water demand increases, with Microsoft reporting a 34\% increase in global water consumption between 2021 and 2022, topping 1.7 billion gallons, while Google observed a 20\% uptick in the same period~\cite{google_esg2024,microsoft_esg2024}. Other studies have sought to estimate water usage at the level of individual AI models, with one paper suggesting that 10–50 queries on GPT-3 consumes around half a liter of water~\cite{li2023making}. By 2050, 50\% of the world’s population is projected to live in a region affected by water scarcity~\cite{boretti2019reassessing}, but the full impacts of data center water usage is unknown as ``the entire data centre industry suffers from a lack of transparency''~\cite{mytton2021data}.

\paragraph{The rare earth minerals needed to produce computing hardware are mined in unsustainable and opaque ways.} Metals such as cobalt, lithium, coltan, gallium, copper, tungsten, and germanium are required for AI hardware and infrastructure. Individual studies have looked at the impacts of mining lithium cobalt, copper, and rare earth for consumer devices (smartphones, tablets) to the GPUs and TPUs powering large-scale AI model training and addressed the resulting environmental damage and impact on local conflict and war. One study specifically addressed mining on indigenous lands, with 54\% of technology-critical materials drawn from what they describe as Indigenous or peasant territory, while 62\% were extracted from drought-prone zones~\cite{owen2023energy}. The mining of these metals also comes with a cost to the environment, given the tonnes of earth that have to be mined, the radiation produced, and the toxic waste created~\cite{dauvergne2022artificial}. Taiwan Semiconductor Manufacturing Company (TSMC), the company that manufactures the GPUs designed by NVIDIA (among other companies), citing the complexity of their supply chains and proprietary information of its customers, also does not provide granular data for their suppliers nor their internal processes (such as chemical purification)~\cite{williams2011environmental}, meaning that it is impossible to carry out a full life cycle analysis of the hundreds of thousands of GPUs that are designed and manufactured each year and used to train and deploy AI models.

\paragraph{AI is responsible for a large amount of greenhouse gas (GHG) emissions.} Despite significant efforts by AI-driven companies to invest in renewable energy and meet net-zero emissions pledges, all evidence indicates that direct GHG emissions due to AI are on the rise. Many AI companies are reporting significantly higher GHG emissions over earlier baselines, likely driven by increases in development and use of generative AI. For example, in their 2024 annual environmental sustainability report (ESG), Google reports a 48\% increase in GHG emissions since 2019 which they attribute primarily to ``increases in data
center energy consumption''~\cite{google_esg2024}, Baidu reports 32.6\% increase in GHG emissions over 2021 citing ``rapid development of LLMs'' posing ``severe challenges'' to their development of green data centers~\cite{baiduesg_2022}, and Microsoft cites a 29.1\% increase since 2020 ``as [they] continue to invest in the infrastructure needed to advance new technologies.''~\cite{microsoft_esg2024}. In fact, one recent analysis found that the real GHG footprints of major tech company data centers can exceed reported values by over 600\%~\cite{obrien2024data}. In response, AI-driven technology firms are increasingly looking to nuclear energy as a lower-carbon alternative to fossil-fuel energy generation that can still provide 24/7 power to datacenters, both by re-commissioning existing infrastructure, such as the Three Mile Island and Susquehanna nuclear power generation facilities in the U.S. state of Pennsylvania~\cite{crownhart2024}, and exploring future development of more modern small modular reactors (SMRs)~\cite{terrell2024}. In addition to the regulatory and security challenges of increasing nuclear power generation capacity, nuclear power presents a distinct set of environmental harms related to disposal of nuclear waste, and increased water consumption required for cooling~\cite{pruavualie2018nuclear}, as well as operational bottlenecks that complicate its widespread adoption~\cite{ojovan2022approaches}. 

\paragraph{AI makes oil and gas extraction processes even more harmful to the environment.}  AI technologies are currently extensively used in oil and gas exploration and drilling, significantly improving efficiency and increasing yield~\cite{sircar2021application,tariq2021systematic}. These \textit{enabled emissions} are difficult to quantify due to lack of corporate transparency and reporting, but they have been attracting scrutiny in recent years, given the environmental harm that they perpetuate. For instance, a coalition of Microsoft employees estimated that a single deal the company struck with Exxon Mobil that uses AI to expand oil and gas production in Texas and New Mexico by 50,000 barrels of oil per day could add up to 640 percent more carbon emissions compared to the company’s carbon removal targets for the year~\cite{grist2024}, yet these numbers were not included in the company's carbon accounting and reporting efforts~\cite{grist2020}. Other firms have been equally opaque about their AI-driven enabling of the oil and gas exploration, but a Greenpeace report from 2020 ties all of the major technology companies to the sector~\cite{oilinthecloud}.

\paragraph{AI contributes to the ballooning issue of electronic waste.} AI's expanding operational footprint also contributes to electronic waste (e-waste), which is now the fastest-growing segment of solid waste worldwide, reaching 62 million tonnes in 2022. Ongoing reports have pointed to the urgent need to find ways to reduce the amount of waste and to reuse these materials more effectively given the status of e-waste as both an environmental and health hazard~\cite{robinson2009waste}. The UN’s Global E-Waste Monitor 2024 showed that about 22\% of e-waste has been shown to be formally collected and recycled, with global generation of electronic waste rising five times faster than e-waste recycling~\cite{balde2024global}. The remainder ends up dumped in landfills, often in developing countries, where researchers assess both how hazardous substances like mercury, arsenic, and lead leach into local ecosystems and how they impact public health. High turnover in AI hardware is accelerating e-waste output: although GPUs can theoretically last about five years, the push for higher performance is prompting more frequent upgrades -- one recent study estimates that AI will generate an additional 1.2–5 million metric tons of e-waste by 2030~\cite{wang2024waste}.

\subsection{Arguments that AI is a net climate positive} \label{climate-positive}
The argument that AI will ultimately be more beneficial than harmful to the environment is primarily based on the narrative that AI can be sustainably developed and used for applications that directly benefit the environment, or that could accelerate the technological advances needed to address climate change. However, claims around AI's sustainability and significance in enabling new scientific advances remain largely hypothetical, with little explicit evidence or quantitative analysis of how the overall impact of AI applications, which may cause equal or greater harm to the environment or even accelerate climate change, will be ultimately beneficial. We discuss the main assertions supporting AI's positive role with respect to climate change in more detail below. 

\paragraph{AI applications can contribute to climate change mitigation and adaptation.} The positive potential of AI for climate action has generated considerable enthusiasm, and there are many existing and in-progress projects and initiatives that aim to harness this potential into concrete tools and applications~\cite{rolnick2022tackling,vinuesa2020role}. There are several key areas in which this is particularly salient, from generating insights from large quantities of data in different modalities to optimizing complex systems and tools~\cite{clutton2021climate}. Concretely, this can be seen in applications of AI that analyze satellite imagery to identify deforestation~\cite{vorotyntsev2021satellite}, methane leaks~\cite{schuit2023automated}, or even the health of coral reefs from space~\cite{burns2022machine}. AI has also improved the accuracy of weather forecasts, which can help renewable energy grid operators predict the output of solar panels and wind turbines, which can help streamline the transition towards renewable energy sources on a global scale~\cite{sharma2011predicting,krechowicz2022machine} In recent years, the development of AI for accelerating scientific discovery in materials science has become an area of focus, especially in applying generative AI models, which are able to assess or generate potential new compounds with climate-positive applications such as renewable energy and carbon capture~\cite{bengio2023gflownet,merchant2023scaling,microsoft-materials}. Finally, the use of AI in solar geoengineering (i.e. injecting aerosol particles into the atmosphere to reflect sunlight and reduce global warming) has also gathered traction, given AI’s potential to help predict the impacts of geoengineering on existing ecosystems and avoid possible negative side effects~\cite{agrawal2025testing}.

\paragraph{AI's negative climate impact is negated by carbon-free energy and carbon offsets.} AI-driven products and services are the primary contributor to AI's direct environmental impacts, and leading AI companies including Google, Microsoft, Amazon and Meta have committed to achieving net-zero carbon emissions for their business operations, including AI development and use, within the next 5-15 years~\cite{google_esg2024,microsoft_esg2024,meta2021,aws-sustain}. The primary mechanism by which these companies plan to achieve their goal is by increasing the use of renewable energy over that period, alongside compensating for emissions using market-based mechanisms such as power purchase agreements (PPAs), , long-term contracts between energy providers and customers who promise to purchase a certain amount of generated energy over a decades-long horizon, carbon offsets, and investment in carbon-reducing technologies. The high-level idea behind these market-based  mechanisms is that any carbon-producing activity can be negated by an equally carbon-reducing activity, resulting in \emph{net zero} emissions globally. In the absence of significant advances in renewable energy production, storage and distribution (which, like other transformative changes in the climate space, would also require significant investment in infrastructure and navigating corresponding sociopolitical systems), PPAs and carbon offsets will remain necessary to fill the gap. However, offsetting was only ever meant to serve as a temporary stop-gap to help reduce emissions in the short term, and does not represent a viable replacement to reducing actual emissions. Fundamental limitations to carbon offsetting are: (1) the difficulty of proving \textit{additionality}, or that the carbon-reducing activity would not have happened regardless of investment or purchase of the offset~\cite{campbell2018barriers,hahn2013understanding}, and (2) that offsetting rarely mitigates localized impacts to the communities where emissions or other environmental degradation is occurring. Increasingly, companies such as Intel and TSMC are reporting annual water withdrawal and consumption as well, and compensating with market-based water offsets which are subject to the same challenges~\cite{woodward2016additionality}.


\paragraph{Increases in efficiency will negate growth in AI resource consumption.}

Another common viewpoint is that the direct environmental impacts of AI will diminish over time due to increases in hardware, software and algorithmic efficiency. \citet{patterson2022carbon} argue that ``the carbon footprint of machine learning training will plateau, then shrink'' thanks to continued innovations in machine learning models, specialized hardware platforms, data center efficiency, scheduling and use patterns, which will reduce overall energy use and emissions. The claim that increasing AI efficiency will lead to an overall reduction in AI's resource use is a clear example of why a deeper engagement with indirect effects is needed in the work on  AI and climate change. 

Similarly to Jevons' Paradox, just because an AI model becomes more efficient, that does not imply that overall AI resource consumption will decrease, and in fact the inverse effect is highly plausible. However, as \citet{KOOMEY20211625} cogently argue, this is not the first time the alarm has been raised about rising energy use due to technology, and that similar projections made in the dot-com boom of the early 2000s failed to materialize. They cite poor data availability, flawed methodology, and inaccurate reporting as causes for inaccurate projections in data center energy use that ultimately did not take into account significant improvements in data center efficiency. Similar uncertainty is clouding the prediction of AI's energy use, and we are well aligned with \citet{KOOMEY20211625}, \citet{doi:10.1126/science.aba3758} and others in calling for more granular transparent data from technology firms and service providers, and more rigorous analysis of available data. This does not imply, however, that AI will necessarily follow the same trend as past technological advances. Further, although data center energy use is a relatively accessible statistic for approximating the environmental impacts of ICT, it represents only small portion of the diverse direct and indirect negative environmental impacts arising from AI, which extend far beyond GHG emissions due to the energy required to develop and use AI models.

\paragraph{There is no reason to focus on AI over other technological advances or sectors.} Some would argue that there is no reason to raise concerns around the current or future energy use (and corresponding environmental impacts) of AI specifically, as compared to any other way that energy might be used, such as ``for watching television, microwaving popcorn, [or] powering lights'' \citep{castro2024rethinking}. This line of thought posits that AI will be subject to the same market pressures, such as energy prices, as any other use case, and as a result will benefit from the same innovations, such as a market transitions to renewable energy. This oversimplification ignores the reality that (1) AI is already responsible for non-trivial negative environmental externalities and (2) we can, and likely should, regulate certain uses of AI as appropriate/inappropriate or necessary/unnecessary, and that this does require breaking down the monolith of ``AI'' into different use cases and corresponding judgments of potential utility. For example, under a limited energy or emissions budget, we might selectively incentivize uses of AI that contribute towards the UN's sustainable development goals~\cite{unsdgs2015} or mitigate national security concerns, over e.g. generating personalized ads for social media (as we discussion in Section~\ref{sec:indirect}). Finally, AI does differ substantially from other energy sinks in its potential to engender vast transformations in the economy and society, similar to how the advent of the Internet has undeniably changed the nature of work, education, and social interactions. 
Purely economic incentives have obviously failed to align well with environmental sustainability in the past, and AI is no different; this does not mean that we should not strive to do better with this new technology.

While debates over AI's role in climate change and sustainability have become increasingly polarized, both sides have tended to focus only on the direct impacts of this technology--- positive and negative. But direct effects are not the whole story. As we show in the next section, the integration of AI into tools and systems reshapes social structures and influences human behavior, which ultimately has complex environmental consequences. 

\section{Indirect Impacts and Rebound Effects} \label{sec:indirect}

 In economics and lifecycle assessment, direct impacts, such as those described in the previous section, are those engendered by the product during its lifecycle, whereas indirect (or second order) impacts refer to systemic responses to the development of the product in terms of behavioral or structural changes which affect on other processes, structures and lifestyles \cite{williams2011environmental,coroamua2020methodology,rivera2014including}. Indirect impacts also include rebound effects, which occur when the improvement of one aspect of a product results in unintended negative consequences due to increased adoption, usage, and workloads~\cite{binswanger2001technological}. These impacts inevitably come with consequences to the environment due to the increased usage, or a redistribution in the usage, of natural resources. Given that AI pervades many different areas of society and economic sectors, it is more difficult to enumerate all of the possible indirect impacts and rebound effects that it can have, which would involve considering the different interactions at play within every sector and between them~\cite{coroamua2020skill,horner2016known}. Nonetheless, our aim in this section is to propose a way of thinking about the indirect environmental impacts of AI that can help inform discussions around AI’s environmental costs and benefits to make them more complete than those described in Section~\ref{sec:direct}. In Table \ref{tab:examples}, we take elements from the qualitative taxonomy for second order environmental effects of ICT proposed by Börjesson, Rivera et al.~\cite{rivera2014including} as well as the initial work on the indirect impacts of AI by Kaack et al.~\cite{kaack2022aligning}. We build upon and adapt both of these approaches to get a better idea of the different types of indirect environmental impacts of AI technologies and, to the extent possible, propose approaches to track and mitigate them to ensure that given interventions do not exacerbate the very things that they were meant to improve. 

\begin{table*}[h!]
\begin{tabular}{lp{10cm}}
\textbf{Indirect Effect}  & \textbf{Examples for AI's environmental impacts}\\ \toprule
Substitution impacts    & \begin{tabular}[c]{@{}l@{}}Art replaced by AI-generated imagery\\ Encyclopedias and books replaced by  AI-generated content \end{tabular} \\ \midrule
Space rebound effects  & Datacenters are getting bigger while devices are getting smaller \\ \midrule
Scale effects & AI models growing in size and complexity\\ \midrule 
Direct economic rebound effects   & AI hardware getting more efficient, yet datacenter energy usage is rising  \\ \midrule
Indirect economic rebound effects &  More consumer devices (speakers, appliances, etc.) that allow users \\ to interact with AI systems  \\ \midrule
Economy-wide rebound effects  & Power purchase agreements from AI compute providers enable global \\ renewable energy capacity to be expanded \\ \midrule
Induction impacts  & Targeted ads that use AI to induce more consumption \\ \midrule
Time rebound effects & \begin{tabular}[c]{@{}l@{}}AI-optimized navigation\\ Robots doing household chores\end{tabular}\\                      \bottomrule               
\end{tabular}
\caption{Examples of AI's indirect impacts and rebound effects, expanding on Börjesson Rivera et al~\cite{rivera2014including}. \label{tab:examples}}
\end{table*}

We further describe each type of effect below, organizing AI’s rebound effects into three themes: material objects and physical spaces (\S\ref{subsec:material}), micro- and macro-economic processes (\S\ref{subsec:economic}), and society and human behavior (\S\ref{subsec:societal}). We provide examples of each, describing the complex ways in which AI-enabled tools and services change existing structures, with unintended ripple effects on the environment.

\subsection{Material Rebound Effects} \label{subsec:material}

AI is impacting the distribution of objects and physical space by changing the way products are made and distributed, and the functionalities that they need to have in order to allow us to interact with AI tools and systems. This has material rebound effects across complex global supply chains, which inexorably changes the way natural resources are exploited, transported, and combined. 

For instance, \textbf{substitution impacts} arise when one product or service replaces another, rendering digital something that was previously analog (i.e. dematerializing it). Recent examples of this include online streaming which has replaced VHS cassettes, vinyl records, CDs and DVDs, as well as e-readers substituting print books and magazines~\cite{hendrickson2010environmental}. This can come with both positive and negative impacts in terms of sustainability: Increased capacity and efficiency of new technologies (e.g. many different movies and TV shows being available on a single streaming platform) can be invalidated by the increased materials needed to support the underlying infrastructure needed to host and deliver these technologies~\cite{rieger2021does}. For instance, a life cycle assessment (LCA), which evaluates the environmental impacts of an artifact arising throughout its existence (typically including disposal), has been performed comparing print books to e-readers, finding that 115 books would produce the same amount of CO$_2$ as a single Amazon Kindle device~\cite{dowd2012kindle,lcaebook}. In the case of AI, many previously material tools and products are being substituted by digital, AI-enabled systems, such as paper maps having been replaced by digital navigation for routine travel. Substitution impacts may occur at varying speeds and may be more gradual. For instance, only a fraction of illustrations such as photographs and artworks have, as of yet, been replaced by AI-generated images, and combinations of AI tools and analog dictionaries are used for translation and writing tasks. While these substitution impacts are likely environmentally-positive (given that it is no longer necessary to manufacture the equipment and supplies needed to carry out the original task), the environmental impacts of developing and deploying these AI-enabled tools remains under-explored and direct comparisons are seldom performed, or even possible using status quo data and methodology (see Section \ref{sec:tracking}). 

Another type of material rebound effect, referred to as the \textbf{scale effect}, occurs when large-scale production or usage of a product has a lesser environmental impact, therefore reducing some of the environmental impacts incurred. For instance, purchasing raw materials in bulk or wholesale lowers overall production costs for businesses, and manufacturing on a larger scale allows a more efficient use of products that would otherwise be wasted~\cite{xue2015computational}. Scaling is a core part of AI research and practice, and the promise of ``methods that continue to scale with increased computation even as the available computation becomes very great''~\cite{sutton2019bitter} has become a core tenet in the field. This is often guided by so-called \textit{scaling laws} that predict the optimal model size and number of training steps based on data availability, incentivizing the development and deployment of increasingly large models as more data becomes available~\cite{hestness2017deep}. At a more granular level, hardware optimizations such as parallelization, do in fact support these economies of scale; e.g. batched inference on a GPU scales sub-linearly in the number of examples, allowing multiple queries to be performed at once at only a marginally higher cost~\cite{silfa2022batch}. In fact, batching, caching and other optimization techniques are routinely used as a means to improve the scaling of AI, allowing more users to interact with ever more powerful technologies~\cite{ramirez2023cache,zhu2024towards}. However, it remains unclear to what extent this pursuit of bigger models requiring more computation is counteracted by optimization approaches that are used in parallel, and the consequences this might have on power grids and supply chains worldwide.

These kinds of shifts in the way objects and processes operate can, in turn, come with \textbf{space rebound effects} that change the way physical space is used with the introduction of a new product. For instance, new approaches such as e-commerce have manifested changes in the physical space occupied by brick-and-mortar stores and warehouses, resulting in smaller stores and bigger warehouses~\cite{tiwari2011environmental}. This, in turn, has impacts on supply chain dynamics, redistributing the environmental footprint of operations (e.g. heating and cooling), logistics and transportation ~\cite{girod2011consumption}. Similarly, the proliferation of videoconferencing platforms and the increased accessibility of high-speed internet, which facilitate remote work, also comes with space rebound effects, with more people working from home, requiring home offices (and therefore bigger homes), and less need for dedicated office space in commercial buildings, with further consequences in urban planning, mobility, and building maintenance ~\cite{hostettler2022potential}. Finally, the size of digital devices has been steadily shrinking over the last decade. For instance, the average mass of a mobile phone shrank by half in the last decade~\cite{hilty2006rebound}, whereas the average size of data centers has doubled~\cite{datacentersize}. With the advent of user-facing AI tools and services running on shrinking mobile devices, the configuration of data centers themselves has shifted from a distributed network of smaller, in-house server rooms to warehouse-sized hyperscale data centers that have tens of thousands of servers in one place. These large, monolithic concentrations of servers enable the large-scale training of AI models, since it allows the high-speed interconnection of thousands of compute nodes serving as a single supercomputer~\cite{hoefler2022convergence}. While hyperscale data centers typically achieve the highest efficiency (measured as PUE), their resource consumption intensity, further magnified by their exceptionally high geographic concentration \citep{IEA2024}, can put undue strain on local infrastructure. For example, Loudoun County in the U.S. State of Virginia has become known as ``Data Center Alley,'' hosting 25 million square feet of data centers that use a quarter of the state’s energy and put extreme strain on its energy supply~\cite{observer2024}. 

 \subsection{Economic Rebound Effects} \label{subsec:economic}

As seen in Section~\ref{sec:direct}, AI is increasingly integrated into economic systems of profit and production, making existing tools more powerful and creating new ones. But these integrations can engender macroeconomic ripple effects across the technology sector as well as adjacent ones, changing structures and processes, as well as transforming whole industries. 

\textbf{Direct economic rebound effects} refer to situations in which the improved efficiency of a product decreases its price, therefore leading to an increase in its consumption. In the centuries since the concept of Jevons' paradox was developed, similar rebound effects have been observed with respect to energy~\cite{giampietro2018unraveling} and water~\cite{dumont2013rebound}, as well as in domains such as road travel (where improvements to roads have been found to result in increased congestion~\cite{thomson1972methods}) and agriculture (where increasing the yield of a crop makes it more profitable to grow it, thereby increasing land use overall~\cite{barbieri2019farming}). They are also the most common type of indirect effects that are discussed in conjunction with many technological innovations (see \cite{kaack2022aligning,willenbacher2021rebound}).
AI is no exception to direct economic rebound effects. While efficiency improvements are being made to the hardware used for training and deploying AI models~\cite{nvidiaefficiency,patterson2022,baiduesg_2022}, NVIDIA shipped 3.7 million GPUs in 2024 (more than a million more units than in 2023) due to increased demand, despite these improvements in efficiency~\cite{ShahNvidia2024}. The data centers that host this hardware are also becoming more efficient, with the average PUE (Power Usage Efficiency) dropping steadily in recent years~\cite{DeepmindPUE,sharma2015analyzing}; however, the energy use and environmental impacts of these data centers has been rising (as described in Section~\ref{sec:direct}).

\textbf{Indirect economic rebound effects} are similar to the direct effects described above, but instead of improved efficiency increasing the usage of the same product, it affects a different product. This kind of rebound effect is also known as a real income effect because the reduced price of one product means that consumers have more income available to spend on other products and services~\cite{hotte2023technology}. For instance, money saved from more fuel-efficient vehicles can be spent on air travel or consumer products~\cite{reimers2021indirect,sorrell2007rebound}. For AI, indirect economic rebound effects can be observed for consumer electronics and ``smart'' devices such as speakers, microwaves and refrigerators. While the primary function of these devices is not AI-driven, they can use AI to propose recipes, answer questions and curate playlists. As more and more AI-enabled consumer tools are developed, the pressure to upgrade existing devices to benefit from these tools increases as well. However, these upgrades come with a price in terms of the natural resources required to manufacture them, as well as to process user data and respond to user queries in real-time~\cite{itten2020digital}. For example, when new functionalities are released on smartphones, they often require purchasing the most recent versions of phones to use them, as was the case with the recent launch of Apple Intelligence, which runs only on the most recent generations of iPhones~\cite{apple2024}, as well as running most of its processing in the cloud, rather than on-device~\cite{eclectic2024}. 

On a more macro level still, \textbf{economy-wide rebound effects} occur when an innovation provokes far-reaching changes in the production and use of other goods, producing flow-on effects in the economy at large. For instance, improvements in global energy efficiency and fuel use have enabled the development of new economies on a global scale, allowing the creation of new industries~\cite{ozsoy2024energy}. These types of rebound effects can be particularly large for so-called ``general-purpose technologies'' such as electricity, the steam engine and the Internet, since these can have global spillover effects on industries and labor markets~\cite{sorrell2009jevons,sorrell2007rebound}. Estimating the scope of economy-wide rebound effects has historically been difficult~\cite{stern2020large}, especially for technologies~\cite{gossart2015rebound}. One way in which AI is concretely impacting the economy at large is via PPAs, as mentioned in Section~\ref{climate-positive}. In 2020, Amazon, Microsoft, Meta, and Google alone accounted for almost 30\% of all PPAs purchased by corporations worldwide~\cite{varro55}, changing the scope and extent of the mechanism as a whole. Renewable energy PPAs mitigate the risk of investment in renewable energy infrastructure, thereby supporting its development~\cite{tantau2022role}. Further, while a portion of the generated renewable energy is transmitted to purchasers in practice, the remainder is added to the energy grid in that region, allowing other users to benefit from its availability. It has also been proposed that foundation models are general purpose technologies with corresponding potential for broad impacts to the economy and labor markets~\cite{eloundou2023gpts}. Analyzing the impact of these changes in the future will be necessary to validate this claim (we discuss approaches to tracking AI's indirect impacts in Section~\ref{sec:tracking}).

\subsection{Societal and Behavioral Rebound Effects} \label{subsec:societal}

As users of technology and citizens of society, our behaviors are shaped and affected by the technologies we use. This has increasingly become the case for AI as it becomes more intertwined in daily activities such as shopping, travel, household chores, and the workplace. In this section, we describe ways in which AI can impact human behaviors and how those changes, can have broader environmental impacts in turn. 

\textbf{Induction effects} occur when the savings in resource consumption (e.g. energy or water) gained through improved efficiency are exceeded by the increased consumption and thereby production of either the materials or the final products. For instance, the advent of fast fashion and mass-produced furniture have reduced the quantity of natural resources, such as wood or cotton, needed to produce a single piece of clothing or furniture. However, the amount of plastics, petrochemicals and waste that are engendered by the increased accessibility of these items and therefore their increased consumption are damaging to the environment~\cite{niinimaki2020environmental}. In the case of AI, a prime example of this is targeted advertising, one of the most lucrative commercial applications of AI, with revenues reaching \$36 billion in 2024~\cite{aimarketing}. Advertising also constitutes the main source of revenue for many technology companies such as Google and Meta~\cite{globaladvertising}; Amazon’s AI-powered product recommendation tool generates almost a third of its annual sales~\cite{mackenzie2013retailers}.  Although it is difficult to obtain quantitative data on the impact of AI on e-commerce, the details of which are largely considered sensitive trade secrets, the corresponding machine learning approaches of  recommendation and ranking are active topics of research in the AI community, powered by an enormous quantity of user data being continuously gathered through interactions with digital platforms~\cite{chen2009factor,dominowska2008first,choi2020identifying}. The rapid expansion of online advertising its direct and indirect environmental impacts have been emphasized in research papers and scientific reports over the last years~\cite{hartmann2023perspectives,planettracker}. The contribution of AI to this expansion is worth considering when measuring the performance of new algorithms and approaches~\cite{dauvergne2022artificial}, especially given the net-zero and emission reduction goals of many AI-driven companies whose revenue models depend on it~\cite{willenbacher2021rebound}. 

\textbf{Time rebound effects} occur when an innovation changes consumers' use of their time, which then frees up (or removes) time for other activities that they carry out. For example, if a vacuum cleaner reduces the time needed to clean the floors in a house, it can free up time for leisure activities such as reading or sports. Measuring the sustainability benefits of these time savings is far from straightforward, because what we do with this newly-freed time can have negative environmental impacts as well, e.g. if time otherwise spent doing chores is now spent shopping or traveling, this can increase one's overall carbon footprint~\cite{buhl2016work}. AI is undeniably impacting how time is spent at work and at home: many activities have already been streamlined using AI over the last decade, such as AI-enabled navigation that avoids traffic congestion, automated license plate recognition and tolling on highways, and robotic vacuums. Initial work on the environmental consequences of these shifts has proposed that AI-enabled automation in household tasks and in the workplace actually leads to greater resource consumption and higher negative environmental impacts overall~\cite{ertel2024rebound}. This contradicts research carried out on AI-assisted navigation specifically, which has been shown to reduce emissions by an average of 3.4\%, based on data gathered by Google on users of Google Maps~\cite{googlemaps}. This discrepancy highlights the difficulty of making far-reaching claims regarding the sustainability impacts of a given innovation without considering the spillover effects and behavioral changes that it may engender more broadly.

\textbf{Indirect policy effects.} In addition to societal and behavioral forces, there are also indirect geopolitical and policy effects. International policy interventions, particularly those aiming to bolster national self-sufficiency from global supply chains, can provoke unintended increases in environmental burdens through mineral demand amplification, industrial deregulation, and competitive overproduction. The dynamics of Koomey’s Law, by which the energy efficiency of computing has historically doubled approximately every 1.57 years, can also be affected by geopolitical disputes and trade wars. If chip sales are restricted, for example, nations have to make do with less powerful and less energy-efficient semiconductors. As national strategies to monopolize or re-shore semiconductor capacity develop, this can further accelerate national rivalries and result in the replication of systems and further resource-intensive scaling of model training infrastructure.
The United States CHIPS and Science Act (2022) is one example. This legislation aims to re-shore semiconductor fabrication. While its primary goal is to reduce dependency on Chinese manufacturing, it has the potential to intensify domestic resource demands. Semiconductor fabrication facilities (known as \textit{fabs}) impose massive environmental burdens, consuming large quantities of energy, water and specialized chemicals while generating substantial carbon emissions (Villard, 2015). The domestic expansion of fabs in the United States shifts rather than reduces these environmental impacts, potentially introducing new ecological pressures in regions with water scarcity issues like Arizona, where several new facilities were planned. Meanwhile the ongoing resource pressures continue in China and Taiwan.
Moreover, the race to decouple US semiconductor supply chains from geopolitical adversaries has led to intensified extraction of rare-earth elements, lithium, cobalt, and high-purity quartz. This geopolitical reorganization of material flows extends the environmental footprint of AI beyond traditional carbon accounting. China restricting the export of several rare earth elements has meant that mines in Ukraine and Australia are already under expanded pressure to meet US demand. The increased competition for critical minerals has both social and ecological impacts, and risks even greater environmental harms in conflict-affected regions.

These indirect policy effects risks are further compounded when environmental deregulation coincides with large-scale investment in AI-related infrastructure. The Trump administration in 2025 has been marked by a comprehensive rollback of environmental protections, including a withdrawal from the Paris agreement, significantly altering the United States' approach to climate and energy policy. Simultaneously, President Trump announced a \$500 billion USD investment in AI infrastructure known as Project Stargate, which aims to significantly increase the number of very large (hyperscale) data centers and bypass environmental impact assessments while offering public subsidies. These policy shifts can lock-in energy and environmental demands that outpace any computational efficiency gains.

This decoupling of AI infrastructure from environmental responsibility illustrates a critical rebound mechanism: technical efficiency improvements, such as high-efficiency datacenter designs or AI model efficiency optimizations, are insufficient when broader regulatory and policy conditions incentivize unchecked growth. By facilitating rapid deployment without ecological constraint, these conditions create an emissions burden with resource demands that scale with model proliferation. In such contexts, even the best-case efficiency improvements under Koomey’s Law may be dwarfed by the sheer increase in volume of compute infrastructure. Then the indirect effects of these policies on the environment are no longer marginal, they become structural.  The recent DeepSeek model is actually an interesting addition to the debate around efficiency and energy because, on the one hand, it was trained using relatively less energy and compute compared to previous generations of language models (due mostly to the export limitations on GPUs to China, which limited the amount of chips that was available for training it), but on the other hand, this was only possible by leveraging synthetic data generated by these models. Also, the model itself is very large --almost 700B parameters for the R1 model -- meaning that deploying it requires access to multiple GPUs with several hundreds of GB of memory, which is significantly more than smaller models of a smaller size, which can contribute to more energy use overall as organizations deploy the model in user-facing applications. Finally, as detailed in the DeepSeek-R1-Zero report~\cite{deepseekai2025}, the model requires much more inference-time computation and energy than previous approaches due to its reasoning abilities. This results in the generation of much longer answers to queries detailing the steps that the model is going through to provide the final response, which leads to more inference-time compute and energy demands.

\subsection{Tracking and Mitigating AI's Rebound Effects} \label{sec:tracking}

The main challenges in measuring and mitigating indirect impacts and rebound effects are their uncertainty and heterogeneity. While tracking direct impacts can largely be achieved by monitoring a finite set of relatively well-defined metrics, such as liters of water consumed or tons of CO$_2$ emitted, rebound effects by definition encompass social, economic, and behavioral impacts across different areas of society~\cite{sovacool2015integrating}. Realizing effective progress in characterizing AI's rebound effects necessitates moving beyond the binary narrative towards more holistic assessment that incorporates a variety of complementary approaches. We discuss some of these approaches below.

\paragraph{Qualitative and quantitative research} Starting with the material rebound effects of digital technologies such as AI, it can be tempting to make comparisons such as the research described in Section~\ref{subsec:material}, which carry out LCA comparisons between e.g. a book and an e-reader. However, we are currently lacking sufficient methodology and data that would allow for meaningful comparison of the environmental impacts of humans versus the same tasks performed by humans alone or assisted by AI. Initial work on this subject has concluded that ``AI systems emit between 130 and 1500 times less CO$_2$e per page of text generated compared to human writers, while AI illustration systems emit between 310 and 2900 times less CO$_2$e per image than their human counterparts,'' however, the authors themselves note that their analyses ``do not account for social impacts such as professional displacement, legality, and rebound effects''~\cite{tomlinson2024carbon}. 
More qualitative and multi-faceted analyses are needed, such as those that take into account aspects like the length of time that content will actually be useful to readers and the way in which the data was collected, can help shed light on the issue, as well as more empirical research to study how human artists and writers use AI technologies and how that impacts the environment.

\paragraph{Transparency} A key challenge to tracking AI's economic rebound effects stems from the severe scarcity of information regarding the nature and extent of AI's actual permeation throughout the economy. The rapid integration of AI into many existing products and services, combined with vague and often broad definitions of AI, has made it difficult to disentangle AI-influenced growth from growth that might have occurred regardless. This challenge is intensified by the perception that details on corporate AI model use represent valuable intellectual property and trade secrets. Gathering more information about specific AI deployment scenarios, i.e. which AI models are actually being deployed, in which sectors, what are they being used for in practice, and what are the impacts of those uses, will become increasingly important as AI becomes established as a mainstream technology. Connecting data on AI model resource consumption to corresponding positive or negative impacts arising from those models is key to understanding tradeoffs and informing decision-making. In the case of induction effects in particular, establishing baselines across industries, for instance documenting the environmental impacts of technologies currently in use, and measuring changes to those baselines as AI is integrated would enable tracking longitudinal trends. Requiring companies using AI in applications such as targeted advertising to indicate this usage, as well as what data was used to choose an ad for a specific product, can help users better understand why they are receiving ads and contribute towards more mindful consumption~\cite{dangelico2017green,minton2012sustainable}.

\paragraph{Regulation} While focusing on efficiency alone has largely been sufficient in the past to curb the environmental impacts of rising compute~\cite{masanet2020recalibrating}, given the astronomical rise in computational (and corresponding energy and natural) resources needed to power AI services resulting from generative AI, complementary efforts are needed to address these impacts~\cite{lbnl2024}. Historically, economic policies that increase the cost of these resources have been shown to help control certain rebound effects~\cite{freire2015energy}. Initiatives such as energy efficiency certifications (e.g. the U.S EPA's Energy Star certification, Energy Labels in the EU) can also reduce resource use by incentivizing the sales and production of more efficient tools and systems, successful for increasing energy efficiency for household appliances~\cite{datta2016analysing}. Related proposals have been introduced for AI models~\cite{luccioni2024light,fischer2022unified}, although the lack of centralized authority for regulating AI (either at a national or international level) has made it difficult so far to operationalize such a proposal. Regulations could also serve an important role in mandating reporting of the necessary data described above, e.g. requiring disclosure of the usage of AI for environmentally harmful applications such as oil and gas exploration, which although highly relevant, is not typically reflected in corporate Environment and Sustainability Governance (ESG) documents.

\paragraph{Efficient devices and distributed compute}
Resource-efficient computing devices that require less natural resources (energy, water, raw materials) to build and use can be seen as an obvious gain in terms of environmental impacts, and different hardware platforms are being developed towards this end as an alternative to GPUs and CPUs, which represent the current status quo of AI computing hardware~\cite{jouppi2017datacenter,jia2019dissecting}. It is worth noting that there is typically a trade-off between resource efficiency and generality in terms of the computational capabilities of the hardware: substantial efficiency gains typically require the hardware to be correspondingly specialized to certain types of computation, such as a specific family of AI models, which can accelerate indirect effects via lock-in, increasing dependence on a narrow set of vendors or products. New approaches to distributed and decentralized computation, in which AI model workloads can be distributed across machines over wide distances rather than running within a single computing cluster or datacenter, can also help distribute environmental burden geographically in terms of resource demand and local impacts. Examples include fully decentralized training~\cite{jaghouar2024intellect} or redirecting training processes depending on which cloud region has the lowest carbon intensity at a given time of day~\cite{dodge2022measuring}. Above and beyond sustainability considerations, adopting more flexible AI hardware and algorithms can help lower the barrier to entry into the field of AI and push back against industry monopolization of computational resources, and corresponding power imbalance that this creates~\cite{abdalla2021grey,abdalla2023elephant}.

\section{Discussion} \label{sec:discussion}

In the context of climate change, artificial intelligence has emerged as a highly polarizing technology. On the one hand, it is championed as a driver of efficiency that could potentially invent new solutions to the climate crisis or reinvent old ones \citep{clutton2021climate, rolnick2022tackling, altman2024intelligence}; on the other hand, it is criticized for its steeply increasing resource demands and carbon footprint \citep{strubell2019energy,li2023making,wang2024waste}. This debate overlooks a significant reality: we still do not have a comprehensive picture of AI’s current environmental impacts on everything from economic systems to individual behaviors. Consequently, the AI field risks simplifying the nuances that must be understood if AI is to be responsibly integrated into environmental policy and practice without exacerbating harms. A more accurate assessment of AI’s environmental outcomes would include a wider range of factors spanning data center operations, supply chains, hardware lifecycles, social behaviors, business incentives, policy commitments, and institutional practices.

As mentioned in Section \ref{sec:indirect}, one under-explored dimension in these debates involves the indirect or rebound effects that AI can generate. While much attention has been given to AI improving productivity and resource efficiency, these gains can result in higher overall consumption due to effects such as Jevons Paradox. This paradox can manifest in several ways: for instance, an AI-driven logistics system might reduce delivery times and fuel usage per vehicle, yet simultaneously encourage more frequent online orders, thus elevating total miles driven; AI-driven targeted advertising can help us more easily find the products that we need based on our clicks and searches, but also push us to make more superfluous purchases. Such systemic shifts in behavior challenge linear expectations that efficiency improvements alone will drive down emissions. Instead, they underscore the need for a detailed, interdisciplinary approach that links AI deployment to broader assessments of environmental, social, and economic feedback loops.

Applying Jevons’ Paradox to technologies like AI has notable conceptual and empirical limitations due to the complexity inherent in technological development and diffusion~\cite{sorrell2008rebound, saunders2009theoretical}. While Jevons’ Paradox anticipates increases in resource consumption following efficiency improvements, its explanatory power diminishes within the broader systemic contexts of widespread AI diffusion. Sorrell~\citeyearpar{sorrell2008rebound} critiques Jevons’ paradox for oversimplifying complex causal relationships among efficiency gains, economic expansion, and individual behavioral responses. This is particularly relevant given AI systems have many interdependencies across diverse economic sectors. Addressing these multifaceted rebound effects requires shifting from simplistic causal assumptions toward more integrated sociotechnical analyses that can include a range of approaches, from detailed empirical studies of technology adoption, user behavior, policy factors, and evolving market dynamics. That would also suggest that truly effective policy responses for AI sustainability would need to combine technological efficiency with institutional reforms and incentive structures specifically designed to decouple model development from unsustainable resource consumption. This approach would not only recognize the conditional validity of Jevons’ Paradox but actively embed sustainability within AI development.

Although the research community has taken initial steps toward quantifying AI’s carbon footprint and water usage, these efforts often focus on direct resource consumption during model training and inference. While such measurements are essential, they are only a partial view. Global supply chains---encompassing the extraction of raw materials, the manufacturing of semiconductor chips, and the disposal of electronic waste---also contribute substantially to AI’s environmental cost, yet these distributed impacts remain notoriously difficult to track \citep{crawford2021atlas,oecd-emissions-supply-chains-2023}. Companies and researchers commonly disclose only a narrow range of environmental metrics. This lack of standardized reporting impedes a full lifecycle assessment and perpetuates an environment in which the indirect burdens of AI remain opaque \citep{luers2024ai}.

Another critical factor is the current market-driven context in which AI operates. At present, the most prominent AI breakthroughs, particularly large language models, function within an economic system that rewards rapid growth and ever-increasing computational power \citep{9563954,doi:10.1126/science.ade2420}. The result is that changing AI development to align with the climate agenda is only possible if it supports extant business incentives---or at least does not limit them. Proposals that transcend profit-driven imperatives, such as limiting the scale of AI (so-called ``digital degrowth''~\cite{selwyn2024digital}) or policy mechanisms such as a carbon tax on data center usage, remain on the fringes of the field. Similarly, few stakeholders are calling for enforced accountability on AI companies to internalize the costs of the environmental damage they cause, including resource depletion, substantial energy consumption, or contributions to e-waste streams. As a result, many of the purported ``solutions'' to climate challenges via AI remain tethered to profit-driven imperatives rather than broader systemic transformations. This structural constraint significantly narrows the scope of AI’s potential as a climate intervention, often leaving only those applications that promise quick returns or minimal disruptions to existing market logics.

This reality highlights a structural barrier: if AI solutions to climate change do not yield near-term financial returns, they may struggle to gain traction in an industry propelled by venture capital, quarterly earnings, and shareholder expectations. Consequently, many of the touted climate positive AI applications, like energy grid optimization or automated water management, risk being overshadowed by more lucrative pursuits that do not necessarily mitigate environmental harm. The underlying problem, then, is not just one of measuring or quantifying impacts more thoroughly within existing market logics --- this simply perpetuates Jevons' paradox. Instead, what is required is a more substantial reimagining of the relationship between AI technologies, business objectives, and ecological imperatives \citep{su12187496}. Genuinely climate-aligned AI strategies might require public policy frameworks that penalize unsustainable practices and reward genuinely carbon-negative deployments of AI, and business models that do not hinge on perpetual growth, in order to ensure that increased AI efficiency does not simply spur more consumption.

Yet, as mentioned in Section~\ref{sec:tracking}, for any of these steps toward meaningful change to materialize, the industry must adopt a far more transparent stance on all the environmental impacts of AI systems and take accountability for the far-reaching impacts of the technologies that it develops and deploys. At present, the scant public information on the carbon footprint of large-scale AI models is frequently derived from academic estimates or limited corporate disclosures rather than comprehensive, standardized reporting. If AI companies truly want to position their technologies as part of the climate solution, they must be forthcoming with granular data on their energy sources, resource consumption, hardware lifecycles, and the end-of-life management of electronic components. In other words, the foundation of any net-positive AI contribution to the environment is a baseline of reliable, detailed data, which has yet to be made widely available. Without this level of transparency, policymaking bodies, researchers, and the public at large are left with partial insights at best, which undermines the capacity to assess AI’s environmental impacts effectively, to design incentives that reward lower-impact AI development, and for individuals to make informed choices with respect to their use of these technologies. Ultimately, the AI field is responsible for knowing the impacts of its own products, and it cannot do so without better data. This information is crucial for performing accurate lifecycle assessments that capture both direct and indirect consequences. Absent such data, the conversation around AI’s climate benefits risks devolving into corporate branding exercises rather than a genuine reckoning with environmental and social responsibilities.

\section{Conclusion}

This paper argues that the AI field needs to adopt a more detailed and nuanced approach to framing, articulating, and addressing AI’s environmental impacts in order to avoid unhelpful polarization. This requires including AI’s direct impacts---mineral supply chain studies, carbon emissions of training large-scale models, energy and water consumption, and e-waste from hardware---as well as mapping the ways AI innovations reshape economic structures and societal practices that, in turn, drive increased resource usage. Such a comprehensive perspective will empower researchers, policymakers, and industry stakeholders to devise strategies that prevent ``tech-solutionism'' from overshadowing the urgent need for systemic change. Greater transparency in reporting energy usage, more robust lifecycle assessment tools, and meaningful industry-wide enforceable standards are examples that would foster much-needed progress.

Ultimately, what is at stake is clear: There is a scientific consensus that the dangers of climate change are extreme, and the effects are already unfolding globally. The need to limit global warming to below 1.5°C underscores the need for transformative change across sectors, and the technology sector is no exception. If AI is deployed without adequate consideration of its direct and indirect effects, it has the potential to deepen inequalities, accelerate resource depletion, and exacerbate the very climate problems it hopes to address. Conversely, if approached with rigorous assessment, transparent reporting, and supportive policy frameworks, AI could serve as a helpful tool in climate adaptation, environmental monitoring, and sustainable planning. Yet we cannot simply hope for the best outcome. The onus is on the AI industry to ensure technology does not contribute to the problem before producing any future solutions. This requires reckoning with AI’s actual impacts, both direct and indirect, measured comprehensively and contextualized socially, economically, and environmentally.

\clearpage
\bibliography{bibliography}
\bibliographystyle{ACM-Reference-Format}

\end{document}